\documentclass[conference]{IEEEtran}
\IEEEoverridecommandlockouts
\usepackage{soul}
\usepackage{cite}
\usepackage{amsmath,amssymb,amsfonts}
\usepackage{algorithmic}
\usepackage{graphicx}
\usepackage{textcomp}
\usepackage{xcolor}
\usepackage{booktabs}
\usepackage{multirow}
\usepackage{authblk}
\usepackage{hyperref}

\def\BibTeX{{\rm B\kern-.05em{\sc i\kern-.025em b}\kern-.08em
    T\kern-.1667em\lower.7ex\hbox{E}\kern-.125emX}}
\begin{document}

\title{DVC-P: Deep Video Compression with Perceptual Optimizations\\ 
}


\author{Saiping~Zhang\textsuperscript{1},
         Marta~Mrak\textsuperscript{2},~\IEEEmembership{Senior Member,~IEEE,}
         Luis~Herranz\textsuperscript{3},
         Marc~G\'orriz~Blanch\textsuperscript{2},
         Shuai~Wan\textsuperscript{4}
         \\and~Fuzheng~Yang\textsuperscript{1}}
         
\affil{\textsuperscript{\tiny 1}\textit{\small State Key Laboratory of Integrated Services Networks, Xidian University, Xi’an, China} \\\textsuperscript{\tiny 2}\textit{\small BBC Research \& Development, The Lighthouse, White City Place, 201 Wood Lane, London, UK}  \\\textsuperscript{\tiny 3}\textit{\small Computer Vision Center, Universitat Autònoma de Barcelona, 08193 Barcelona, Spain} \\\textsuperscript{\tiny 4}\textit{\small School of Electronics and Information, Northwestern Polytechnical University, Xi’an, China}}

\maketitle

\begin{abstract}
Recent years have witnessed the significant development of learning-based video compression methods, which aim at optimizing objective or perceptual quality and bit rates. In this paper, we introduce deep video compression with perceptual optimizations (DVC-P), which aims at increasing perceptual quality of decoded videos. Our proposed DVC-P is based on Deep Video Compression (DVC) network, but improves it with perceptual optimizations. Specifically, a discriminator network and a mixed loss are employed to help our network trade off among distortion, perception and rate. Furthermore, nearest-neighbor interpolation is used to eliminate checkerboard artifacts which can appear in sequences encoded with DVC frameworks. Thanks to these two improvements, the perceptual quality of decoded sequences is improved. Experimental results demonstrate that, compared with the baseline DVC, our proposed method can generate videos with higher perceptual quality achieving 12.27\% reduction in a perceptual BD-rate equivalent, on average.
\end{abstract}


\section{Introduction}
Various video services, taking ultra high-definition videos and panoramic videos as examples, have brought great challenges to video compression methods. In past decades, traditional video coding standards, from H.264/AVC \cite{b2} to H.266/VVC \cite{b3}, have achieved tremendous development in saving bit rates and enhancing quality of decoded videos. These achievements mainly rely on very carefully designed modules in block-based hybrid coding framework. Recently, new approaches based on Deep Neural Networks (DNN) adopted a different strategy. In particular, DNN based video compression methods pay more attention to end-to-end optimization instead of carefully designing a specific module in video compression framework. Although still in a research phase, such strategy has a potential to provide better compression and revolutionize video compression field.

In past few years, a number of deep network designs for video compression have been proposed, achieving promising results in terms the trade off between rate and objective distortion (e.g.  peak signal to noise ratio, PSNR) performance. Lu et al.\cite{b5} firstly designed a deep end-to-end video compression (DVC) model that established a one-to-one correspondence between modules of conventional hybrid video coding framework and their model. Furthermore, to alleviate error propagation and enable coding adaptation to different types of video content, they proposed an improved DVC model\cite{b6}. Lin et al.\cite{b8} employed multiple reference frames to help predict current frame more accurately, yielding less residual.

However, optimizing compression towards improving PSNR does not always improve perceptual quality of decoded videos. Considering optimizing a video compression network towards higher perceptual quality, recently proposed methods deploy Generative Adversarial Networks (GANs). Zhu et al.\cite{b16} employed GAN to remove the spatial redundancy in video frames and improved the performance of intra prediction in video coding process. But only improving intra-coded frames is insufficient for enhancing the performance of the whole decoded video. Veerabadran et al.\cite{b18} presented an adversarial learned video compression model based on a 3D autoencoder, which tends to eliminate blurred results under extreme video compression. However, 3D convolutions are difficult to train because of a large number of parameters, which put a limitation on improving the perceptual quality of decoded videos. 

In this paper, a deep video compression with perceptual optimizations (DVC-P) network is proposed, which aims at optimizing for perceptual quality of decoded videos. The main contributions of our work are summarized as follows:

(1) We optimized DVC with a discriminator network and a mixed loss to enhance perceptual quality of decoded videos.

(2) We eliminated checkerboard artifacts in DVC with nearest-neighbor interpolation, and further improve perceptual quality of decoded videos.

(3) We evaluated performance of the proposed DVC-P in terms of Fréchet video distance (FVD) \cite{b27} which is a metric highly correlated to human visual experience of videos and a BD-rate equivalent. The proposed DVC-P has outperformed DVC \cite{b28} in terms of FVD scores and achieved 12.27\% reduction in a BD-rate equivalent.

\begin{figure}[t]
\centerline{\includegraphics[scale=0.42]{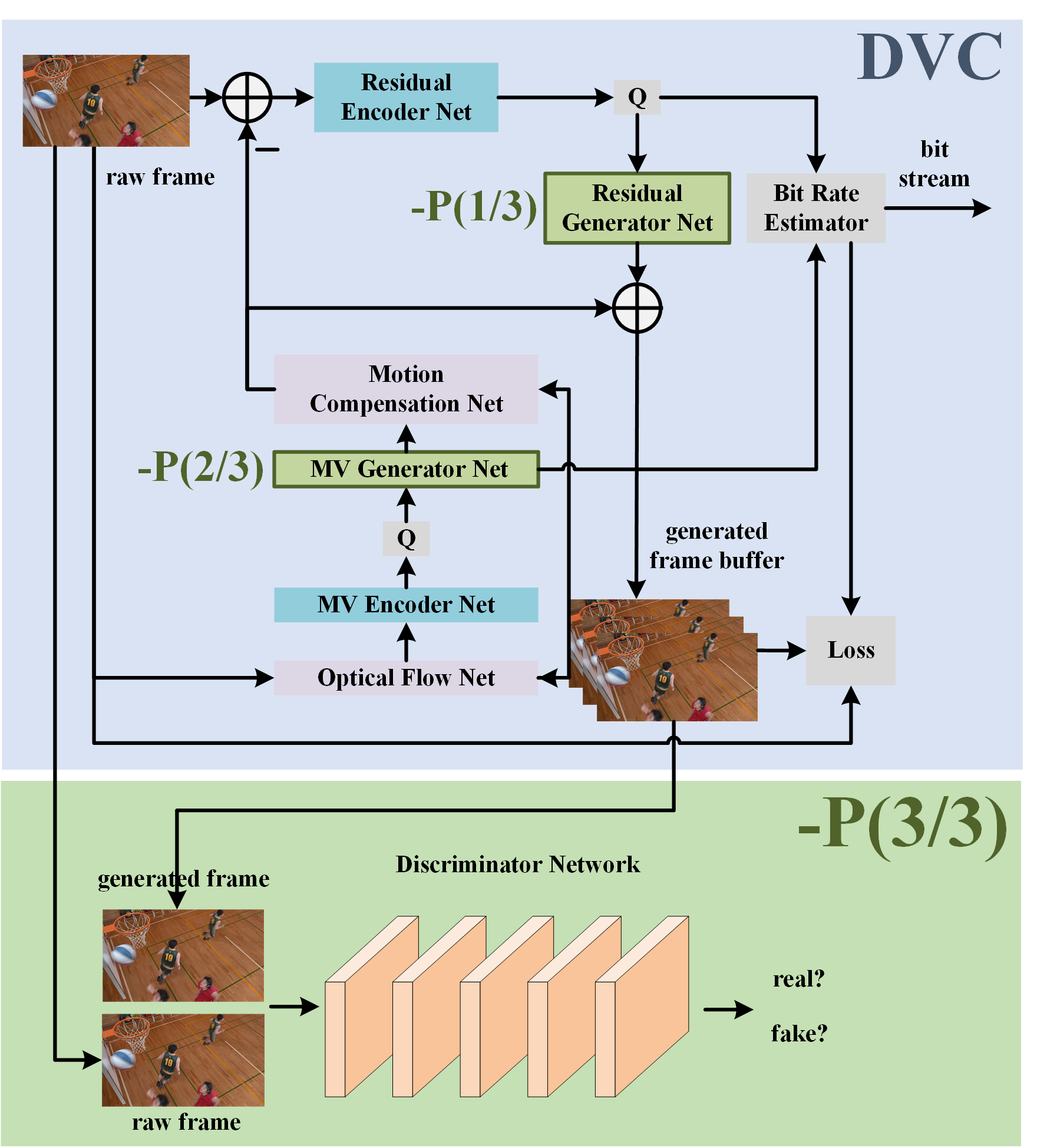}}
\caption{The proposed DVC-P network}
\label{fig.1}
\end{figure}

\section{Proposed Method}

The structure of the DVC-P network is shown in Fig.~\ref{fig.1}, where three proposed improvements are shown in green. ``-P(1/3)" and ``-P(2/3)" modules can enhance synthesis of pixels, and ``-P(3/3)" module can guide generated frames optimized towards real frames.

\subsection{Baseline Deep Video Compression Network}\label{3A}

 The structures of residual encoder network, motion vector (MV) encoder network, optical flow network, motion compensation network and bit rate estimation follow those in DVC network\cite{b5}. Specifically, residual encoder network, which encodes residuals between the raw video frame and reconstructed video frame to bit streams, consists of four convolution layers. Each layer downsamples its input with stride=2. There is a rectifier unit (ReLu) after every convolution except the last one. After quantization, the signal is losslessly processed by entropy coding to form the bit stream. Since both are non-differentiable, during training quantization is replaced by additive uniform noise\cite{b19}, and entropy coding is bypassed, approximating rate by the entropy of the latent representation. As for MV encoder network, its structure follows the same design as the residual encoder network. A pretrained optical flow estimation network\cite{b20} is used to estimate motion between the generated/reference frame and current raw frame. It is fine tuned during the training process. The motion compensation network achieves warp operation and prepares for residual calculation. In terms of estimating the bit rate, the entropy model in\cite{b19} is used to calculate it.

\subsection{Perceptual Optimizations}
\subsubsection{Proposed Generator and Discriminator}\label{3A}
At the encoder side, residuals and MVs are reconstructed using generator networks, with the purpose to generate reference frames for inter coding. Four convolution layers are designed in both generator networks. Instead of using common strided-deconvolution in these generator networks, we use nearest-neighbor interpolation to achieve upsampling and restore the original resolution of signals. Activation function is ReLu. Moreover, in our proposed DVC-P we implement the discriminator from DCGAN\cite{b21} which proposed a set of constraints on the architecture of discriminator networks to make them stable to train in most settings.

\subsubsection{Mixed Loss function}\label{3B}
The total loss of the proposed DVC-P is formulated as the weighted sum of MSE loss, adversarial loss, VGG-based loss and bit rate loss as: 

\begin{equation}
G_{Loss} = \alpha MSE + \beta {{Loss_{adv}}} + \gamma {{Loss_{vgg}} + \omega {{Loss_{R}}}}\label{eq1}
\end{equation}
where $MSE$, $Loss_{adv}$, $Loss_{vgg}$ and $Loss_{R}$ represent MSE loss, adversarial loss, VGG-based loss and bit rate loss, respectively. $\alpha$, $\beta$, $\gamma$ and $\omega$ are the corresponding weights.

Adversarial loss is computed as:
\begin{equation}
Loss_{adv} = {{\rm{{\rm E}}}_{z\sim{{p_z}(z)}}\left[ {{{\left( {D\left( {G\left( z \right)} \right) - 1} \right)}^2}} \right] \label{eq2}}
\end{equation}
where $z$ is the input of the generator. $G\left( {} \right)$ represents the generator, and $D\left( {} \right)$ represents the discriminator. We use the least squares loss function from LSGAN\cite{b23} which solved vanishing gradients problem during the training process.

VGG-based loss is computed as:
\begin{equation}
Loss_{vgg} = {{\left\| {F(x) - F\left( {G\left( z \right)} \right)} \right\|_1} \label{eq3}}
\end{equation}
where $x$ represents the raw frame, and $F\left( {} \right)$ represents the features of the 4th convolution before the 5th max-pooling layer of an ImageNet pretrained VGG-19 network\cite{b24}.

MSE loss is essential for video compression networks to maintain the video content unchanged. On the other hand, adversarial loss can help generators produce decoded videos of higher perceptual quality. Moreover, incorporating VGG-based loss is beneficial to stabilizing the whole training process.

Bit rate loss is computed as:
\begin{equation}
Loss_{R} =  - {\rm{E}}\left[ {{{\log }_2}{P_{\widehat r}}} \right] - {\rm{E}}\left[ {{{\log }_2}{P_{\widehat m}}} \right]
\label{eq4}
\end{equation}
where ${P_{\widehat r}}$ and ${P_{\widehat m}}$ represent the probabilities of residuals and MVs after quantization. 

The loss of discriminator is computed as in LSGAN \cite{b23}:

\begin{equation}
\begin{aligned}
D_{Loss} &= {0.5 \times {\rm{{\rm E}}}_{z\sim{{p_z}(z)}}\left[ {{{\left( {D\left( {G\left( z \right)} \right)} \right)}^2}} \right]}\\
&+ {0.5\times\rm{{\rm E}}}_{x\sim{{p_x}(x)}}\left[ {{{\left( {D\left( x\right)-1} \right)}^2}} \right] \label{eq5}
\end{aligned}
\end{equation}

\subsubsection{Elimination of Checkerboard Artifacts}\label{3A}
 Deconvolution has uneven overlap in two dimensions (i.e., \textit{x} dimension and \textit{y} dimension) when the ``kernel size" is not divisible by the ``stride", which sometimes leads to checkerboard artifacts in the final outputs. An efficient and effective way to solve this issue is upsampling images by nearest-neighbor interpolation (or Bilinear interpolation) and followed by a convolution layer (stride=1)\cite{b22}.  Furthermore, adversarial loss can further help improve visual quality of decoded frames. (see Fig. \ref{fig.2} for the visualization results.)
 
\section{Experimental Results}

\subsection{Dataset}
We use Vimeo-90k\cite{b25} dataset to train our proposed DVC-P. 7 consecutive frames in a video sequence are regarded as a sample and cropped in 256x256 before fed into the network. Frames in the same sample are cropped in the same position, but frames in different samples are cropped randomly. Batch size is 4. For evaluating the performance of our proposed DVC-P, tests are performed on JCT-VC test sequences \cite{b26}.

\subsection{Training Strategy}
Similarly to the baseline DVC in which the framework design consists of various deep models, our proposed DVC-P requires carefully designed joined training strategy. In particular, the training process consists of 700k iterations in total. When $iterations<20k$, only optical flow network, MV encoder network and MV generator network are trained together. When $iterations$ reaches to $20k$, motion compensation network begins to join the training. When $iterations$ reaches to $40k$, residual encoder network and residual generator network also begin their joint training. When $iterations$ reaches to $400k$, the discriminator begins to be optimized. As for loss function, we only use MSE loss when $iteration<20k$, VGG-based loss is added when $iterations$ reaches to 40k. Adversarial loss is added when $iterations$ reaches to 400k. Learning rate is set ${10^{ - 4}}$ during the whole training.

\subsection{Results}
For evaluation of the proposed DVC-P, the following training parameters for Eq. (\ref{eq1}) are used: $\alpha=1$, $\beta=0.1$ and $\gamma=0.04$. Different $\omega$ in Eq.(\ref{eq1}) leads to different rate-distortion-perception trade-off. The GOP size is 10, and the first 100 frames are tested for each sequence. 

\subsubsection{Perceptual Video Quality Metric}
\begin{figure}[htbp]
\centerline{\includegraphics[scale=0.4]{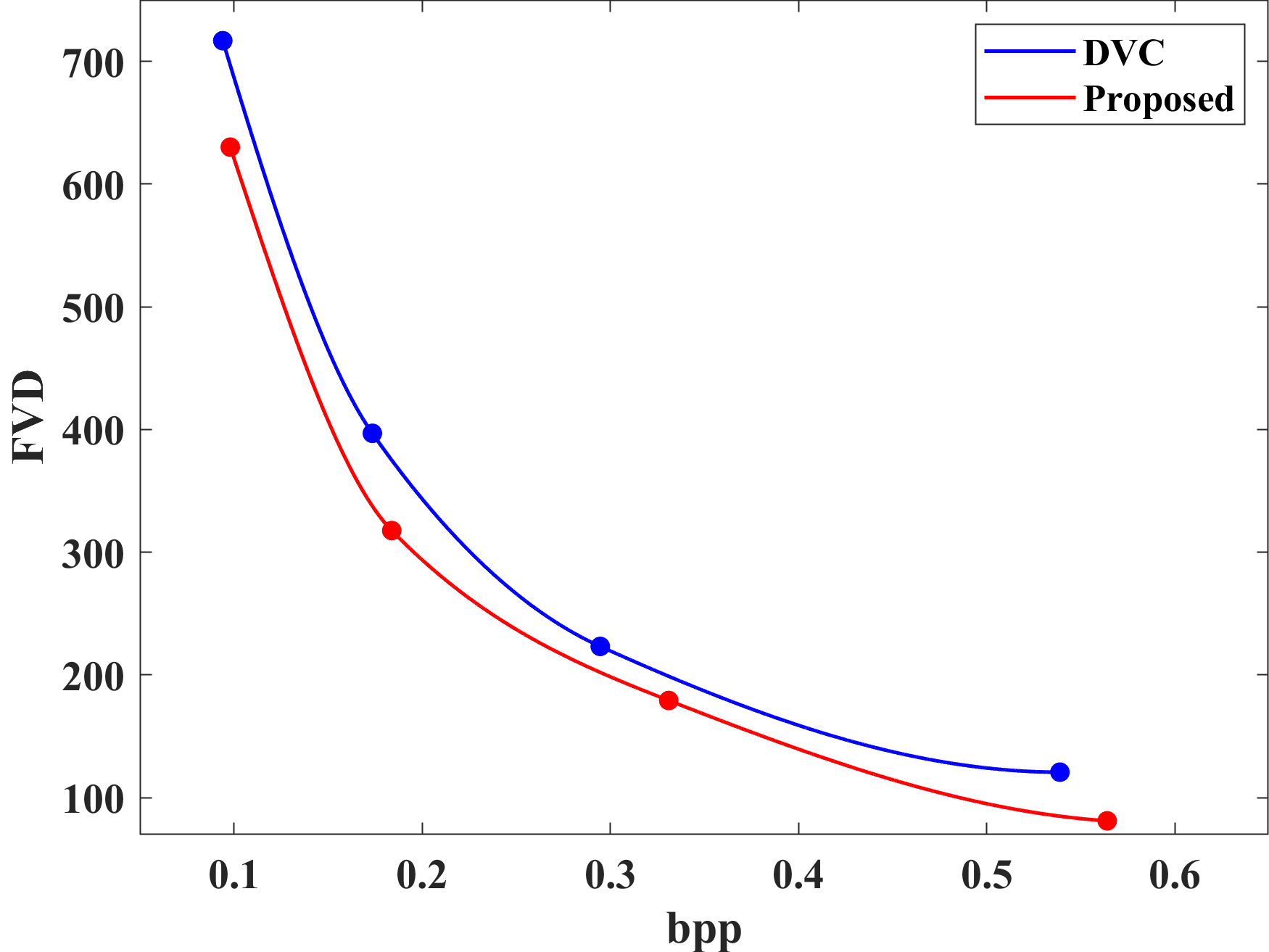}}
\caption{Performance comparison of the proposed DVC-P with DVC for sequence \textit{RaceHorses(class D)}}
\label{fig.4}
\end{figure}

We test perceptual quality of decoded videos by FVD. When setting $\omega={{\rm{1}} \mathord{\left/
 {\vphantom {{\rm{1}} {{\rm{256}}}}} \right.
 \kern-\nulldelimiterspace} {{\rm{256}}}}$ for proposed DVC-P and $\lambda=256$ for DVC ($\lambda$ trades off between distortion and bit rate in DVC. $\lambda=256$ corresponds to QP=37 in DVC), we compute FVD for all sequences at almost the same bit rate, as shown in Table \ref{Table1}. Smaller FVD values correspond to better performance. We also compute a BD-rate equivalent (referred to ``FVD BD-rate") which indicates how much less bit rate the proposed method needs to achieve the same FVD as DVC for the same FVD, over 4 QP points: 22, 27, 32 and 37 (corresponding to $\lambda$ = 2048, 1024, 512 and 256, $\omega$ = ${{\rm{1}} \mathord{\left/
 {\vphantom {{\rm{1}} {{\rm{2048}}}}} \right.
 \kern-\nulldelimiterspace} {{\rm{2048}}}}$, ${{\rm{1}} \mathord{\left/
 {\vphantom {{\rm{1}} {{\rm{1024}}}}} \right.
 \kern-\nulldelimiterspace} {{\rm{1024}}}}$, ${{\rm{1}} \mathord{\left/
 {\vphantom {{\rm{1}} {{\rm{512}}}}} \right.
 \kern-\nulldelimiterspace} {{\rm{512}}}}$ and ${{\rm{1}} \mathord{\left/
 {\vphantom {{\rm{1}} {{\rm{256}}}}} \right.
 \kern-\nulldelimiterspace} {{\rm{256}}}}$), as shown in Table II. Notice that DVC-P performs worse on \textit{BQSquare}. It is because on smaller QPs (22 and 27), where FVD is already low, the bit rate is higher. If we just focus on larger QPs (32 and 37), DVC-P still performs better. In addition, We draw ``FVD-Bit rate'' curves in Fig.\ref{fig.4} to compare the performance at 4 QPs, taking sequence \textit{RaceHorses(class D)} as an example. In general, our proposed method can generate more realistic decoded videos and outperform DVC.

\newcommand{\tabincell}[2]{\begin{tabular}{@{}#1@{}}#2\end{tabular}}  
\setlength\tabcolsep{3.5pt}
\begin{table}[htbp]
\caption{FVD and Bitrate Comparison of Standard Test Sequences at QP=37 ($\lambda=256$ and $\omega={{\rm{1}}\mathord{\left/
 {\vphantom {{\rm{1}} {{\rm{256}}}}} \right.
 \kern-\nulldelimiterspace} {{\rm{256}}}}$)}
\begin{center}
\begin{tabular}{cccccc}
\toprule
\multicolumn{2}{c}{\multirow{2}{*}{\textit{Sequence}}} & \multicolumn{2}{c}{DVC \cite{b5}} & \multicolumn{2}{c}{Proposed} \\ 

& & FVD &Bitrate (bpp)  & FVD & Bitrate (bpp) \\ 
\midrule 
\multirow{2}{*}{A}     & \textit{Traffic}         & 590.02  & 0.041 &458.53 & 0.044\\
                       & \textit{PeopleOnStreet}  & 593.01  & 0.073 &566.25 & 0.074\\
\hline
\multirow{5}{*}{B}     & \textit{Kimono}          & 207.02  & 0.046 &156.05 & 0.054\\
                       & \textit{ParkScene}       & 411.47  & 0.044 &324.74 & 0.047\\
                       & \textit{Cactus}          & 572.01  & 0.050 &453.19 & 0.054\\
                       & \textit{BQTerrace}       & 449.83  & 0.053 &369.08 & 0.055\\
                       & \textit{BasketballDrive} & 552.21  & 0.059 &435.10 & 0.062\\
\hline
\multirow{4}{*}{C}     & \textit{RaceHorses}      & 437.78  & 0.094 &385.52 &0.099\\
                       & \textit{BQMall}          & 566.18  & 0.073 &425.13 &0.076\\
                       & \textit{PartyScene}      & 571.08  & 0.103 &446.83 & 0.104\\
                       & \textit{BasketballDrill} & 674.06  & 0.056 &528.84 & 0.059\\
\hline
\multirow{4}{*}{D}     & \textit{RaceHorses}      & 716.85  & 0.094 &630.10  & 0.098\\
                       & \textit{BQSquare}        & 1007.97 & 0.091 &905.61  & 0.094\\
                       & \textit{BlowingBubbles}  & 811.58  & 0.089 &615.89  & 0.091\\
                       & \textit{BasketballPass}  & 876.21  & 0.062 &623.76  & 0.065\\
\hline
\multirow{3}{*}{E}     & \textit{FourPeople}      & 289.34  & 0.029 &248.15 & 0.031\\
                       & \textit{Johnny}          & 278.51  & 0.021 &234.55 & 0.023\\
                       & \textit{KristenAndSara}  & 213.72  & 0.024 &195.45 & 0.026\\
\hline
\multicolumn{2}{c}{\textit{Average}}              & 545.49  & 0.061 &444.60  & 0.064\\     
\bottomrule 
\end{tabular}
\end{center}
\label{Table1}
\end{table}

\subsubsection{Elimination of Checkerboard Artifacts}
\begin{figure*}[]
\centerline{\includegraphics[scale=0.48]{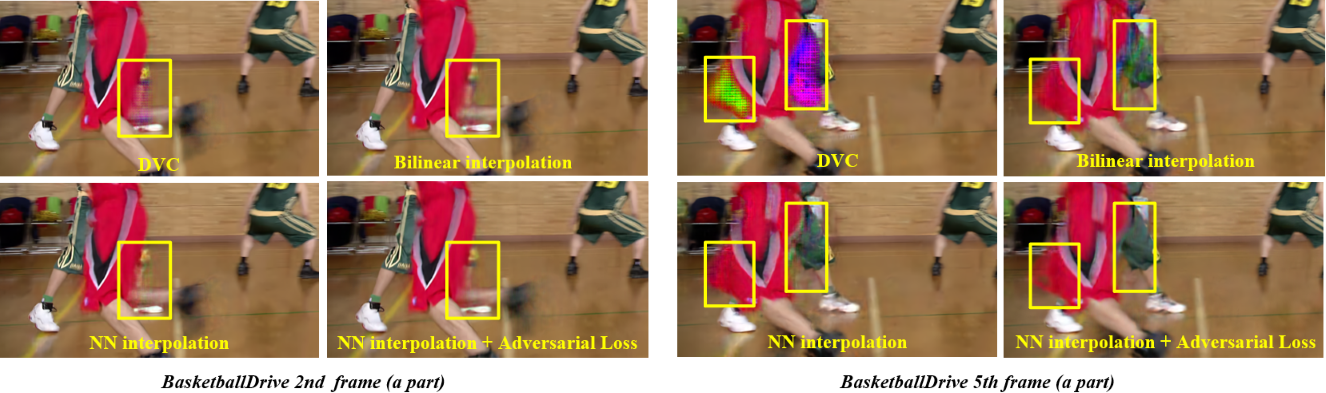}}
\caption{Comparison of elimination of checkerboard artifacts (NN refers to nearest-neighbor)}
\label{fig.2}
\end{figure*}

Deconvolutions tend to bring checkerboard artifacts (sometimes very strong artifacts), which leads to colorful blurs appear in the decoded video \textit{BasketballDrive} compressed with DVC ($\lambda=256$). Nearest-neighbor interpolation and Bilinear interpolation can eliminate this kind of artifacts to some extent. Besides, our proposed GAN ($\omega={{\rm{1}} \mathord{\left/
 {\vphantom {{\rm{1}} {{\rm{256}}}}} \right.
 \kern-\nulldelimiterspace} {{\rm{256}}}}$) with adversarial loss can further improve perceptual quality of this decoded video. Visualization comparison is shown in Fig.\ref{fig.2}. Thanks for nearest-neighbor interpolation and adversarial loss, checkerboard artifacts are eliminated, and perceptual quality is satisfactory. Although checkerboard artifacts only appear in \textit{BasketballDrive}, we test FVD and PSNR values of all sequences to explore the influence of nearest-neighbor and Bilinear interpolation methods, as shown in Table II.

\subsubsection{Visual Comparison}

\begin{figure*}[]
\centerline{\includegraphics[scale=0.50]{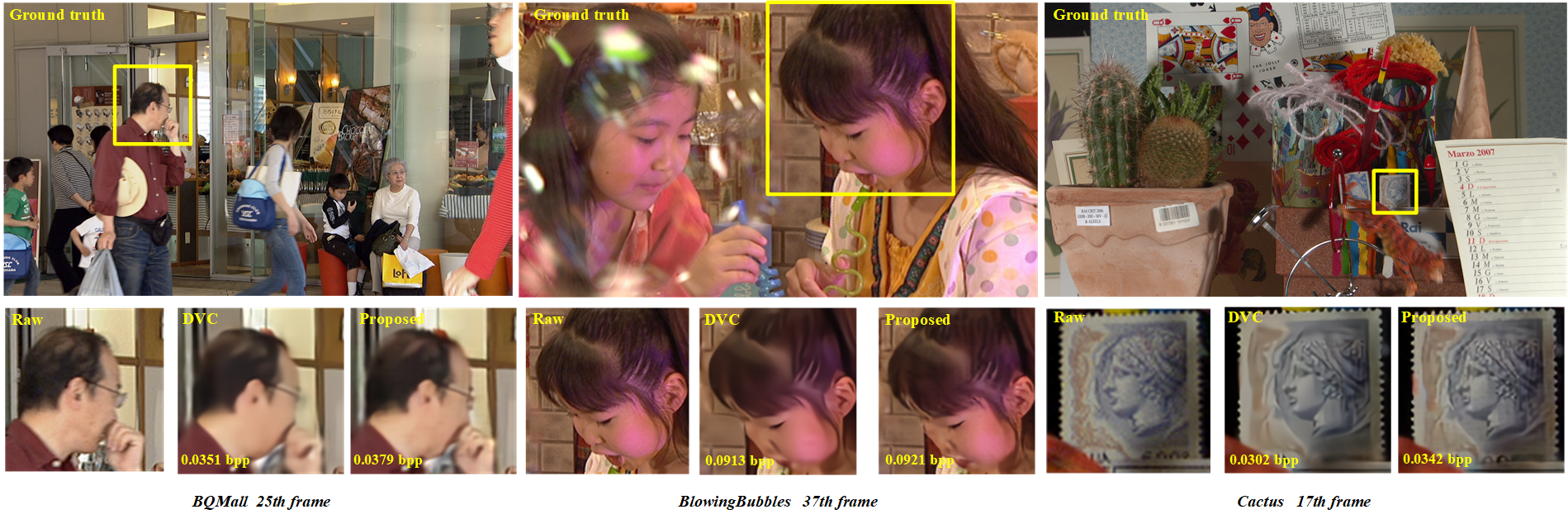}}
\caption{Visual comparison of the proposed DVC-P with DVC (The number of bpp represents the corresponding bit rate of the whole frame)}
\label{fig.3}
\end{figure*}

Visual comparison between  proposed DVC-P and DVC is shown in  Fig.\ref{fig.3} for three randomly selected areas in inter-coded frames. It can be seen that  DVC has more blurred areas, which penalizes the perceptual quality.

\section{Conclusion}
In this paper, we have proposed the DVC-P network aiming at restoring decoded videos in high perceptual quality at the decoder side. By using a discriminator and a mixed loss to guide the whole video compression network to optimize towards generating realistic decoded videos, our proposed DVC-P outperformed DVC in terms of a BD-rate equivalent and visual experience. 

\setlength\tabcolsep{8pt}
\begin{table*}[htbp]
\caption{FVD and PSNR Comparison and a BD-rate Equivalent of Standard Test Sequences (``NN" refers to ``nearest-neighbor") }
\begin{center}
\begin{tabular}{cc|cccccccc|c}
\toprule
\multicolumn{2}{c|}{\multirow{4}{*}{\textbf{Sequence}}}       & \multicolumn{8}{c|}{\multirow{2}{*}{QP = 37}} & QP = \{22, 27, \\
& & & & & & & & & & 32, 37\} \\ \cline{3-11} 
&       & \multicolumn{2}{c}{\textbf{DVC\cite{b5}}} & \multicolumn{2}{c}{\textbf{Bilinear}} & \multicolumn{2}{c}{\textbf{NN}} & \multicolumn{2}{l|}{\textbf{NN+Adversarial Loss}} & \multirow{2}{*}{\textbf{\tabincell{c}{FVD\\BD-rate}}} \\ 
                   &                          & FVD    & PSNR           & FVD    & PSNR          & FVD       & PSNR       & FVD                & PSNR &                   \\
\hline
\multirow{2}{*}{A} & \textit{Traffic}         &590.02  &\textbf{31.96}  &564.03  &31.93  &575.08  &\textbf{31.96} &\textbf{458.53}  &31.82          &\textbf{-16.55\%} \\
                   & \textit{PeopleOnStreet}  &593.01  &\textbf{31.28}  &640.46  &30.47  &609.66  &30.96          & \textbf{566.25}  &31.13          &\textbf{-2.95\%}  \\
\hline
\multirow{5}{*}{B} & \textit{Kimono}          &207.02  &34.62           &211.62  &34.47  &220.56  &34.62          &\textbf{156.05}  &\textbf{34.81} &\textbf{-14.12\%} \\
                   & \textit{ParkScene}       &411.47  &\textbf{31.22}  &407.70  &31.06  &412.63  &31.14          &\textbf{324.74}  &31.17          &\textbf{-13.74\%} \\
                   & \textit{Cactus}          &572.01  &\textbf{30.17}  &618.71  &30.00  &612.05  &30.15          &\textbf{453.19}  &29.98          &\textbf{-26.40\%} \\
                   & \textit{BQTerrace}       &449.83  &\textbf{30.00}  &439.58  &29.90  &449.12  &29.98          &\textbf{369.08}  &29.84          &\textbf{-9.50\%}  \\
                   & \textit{BasketballDrive} &552.21  &28.74           &452.65  &30.34  &448.62  &\textbf{30.78} &\textbf{435.10}  &30.70          &\textbf{-7.03\%}  \\
\hline
\multirow{4}{*}{C} & \textit{RaceHorses}      &437.78  &\textbf{27.75}  &468.58  &27.44  &444.40  &27.65          &\textbf{385.52}  &27.69          &\textbf{-12.46\%} \\
                   & \textit{BQMall}          &566.18  &27.77  &614.02  &27.72   &590.29 &\textbf{27.88}          &\textbf{425.13}  &27.64          &\textbf{-8.57\%}  \\
                   & \textit{PartyScene}      &571.08  &\textbf{26.18}  &600.88  &26.07  &596.08  &\textbf{26.18} &\textbf{446.83}  &25.98          &\textbf{-2.43\%}  \\
                   & \textit{BasketballDrill} &674.06  &\textbf{29.86}  &689.69  &29.58  &666.61  &29.80          &\textbf{528.84}  &29.66          &\textbf{-10.51\%} \\
\hline
\multirow{4}{*}{D} & \textit{RaceHorses}      &716.85  &\textbf{27.54}  &767.19  &27.23  &742.58  &27.50          &\textbf{630.10}  &27.51          &\textbf{-16.30\%} \\
                   & \textit{BQSquare}        &1007.97 &26.97           &1013.62 &26.95  &1028.34 &\textbf{27.08} &\textbf{905.61}  &26.57          &\textbf{3.62\%}   \\
                   & \textit{BlowingBubbles}  &811.58  &\textbf{27.15}  &857.07  &27.02  &815.75  &\textbf{27.15} &\textbf{615.89}  &27.00          &\textbf{-13.49\%} \\
                   & \textit{BasketballPass}  &876.21  &28.84           &806.65  &28.66  &785.32  &\textbf{28.86} &\textbf{623.76}  &28.70          &\textbf{-18.12\%} \\
\hline
\multirow{3}{*}{E} & \textit{FourPeople}      &289.34  &34.48  &279.09  &\textbf{34.55}  &291.06  &34.54          &\textbf{248.15}  &33.92          &\textbf{-20.18\%} \\
                   & \textit{Johnny}          &278.51  &35.77  &275.73  &\textbf{35.87}  &263.00  &35.86          &\textbf{234.55}  &35.38          &\textbf{-15.37\%} \\
                   & \textit{KristenAndSara}  &213.72  &35.20  &208.42  &\textbf{35.36}  &214.92  &35.30          &\textbf{195.45}  &34.65          &\textbf{-16.71\%} \\
\hline
\multicolumn{2}{c|}{\textit{Average}}          &545.49  &30.31  &550.87  &30.26           &542.56  &\textbf{30.41} &\textbf{444.60}  &30.23          &\textbf{-12.27\%} \\         
\bottomrule
\end{tabular}
\end{center}
\end{table*}

\vspace{12pt}
\end{document}